\documentclass[12pt]{article}
\usepackage{amsmath}
\def\s{\sigma}

\def\be{\begin{equation}}
\def\ee{\end{equation}}
\def\arr{\begin{array}{rll}}
\def\ea{\end{array}}
\def\bea{\begin{eqnarray}}
\def\eea{\end{eqnarray}}

\def\N2{$N{=}2$}

\def\tr{{\rm tr}}

\def\>{\rangle}
\def\<{\langle}
\def\+{\dagger}
\def\={\ =\ }

\begin{document}
\renewcommand{\thefootnote}{\fnsymbol{footnote}}
\begin{titlepage}
\setcounter{page}{0}
\begin{flushright}
LMP-TPU--01/12  \\
\end{flushright}
\vskip 1cm
\begin{center}
{\LARGE\bf  Higher rank Killing tensors  }\\
\vskip 0.5cm
{\LARGE\bf  and Calogero model}\\
\vskip 2cm
$
\textrm{\Large Anton Galajinsky \ }
$
\vskip 0.7cm
{\it
Laboratory of Mathematical Physics, Tomsk Polytechnic University, \\
634050 Tomsk, Lenin Ave. 30, Russian Federation} \\
{Email: galajin@mph.phtd.tpu.ru}

\end{center}
\vskip 1cm
\begin{abstract} \noindent
$(n+2)$--dimensional Lorentzian spacetime which admits irreducible Killing tensors of rank up to $n$ is constructed by applying the
Eisenhart lift to the Calogero model.
\end{abstract}

\vspace{0.5cm}

\end{titlepage}

\renewcommand{\thefootnote}{\arabic{footnote}}
\setcounter{footnote}0

\noindent
{\bf 1. Introduction}\\

In Riemannian geometry symmetries of a metric are conventionally described by Killing vectors which generate infinitesimal coordinate
transformations in spacetime. When considering the geodesic equation, each Killing vector field with components $\xi^n(x)$ gives rise to the first integral $\xi^n(x) g_{nm} (x) \frac{d x^m}{d \tau}$, where $g_{nm} (x)$ is a metric. In general, a spacetime may also admit Killing tensors, i.e. totally symmetric tensor fields obeying the equation $\nabla_{(i_1} K_{i_2 \dots i_{n+1} )}=0$, which underlie the first integrals $K_{i_1 \dots i_n} (x) \frac{d x^{i_1}}{d \tau}\dots \frac{d x^{i_n}}{d \tau}$ of the geodesic equation. The existence of the Killing tensors is usually attributed to hidden symmetries of spacetime as there is no coordinate transformation associated to them.

In some instances, the logic can be turned around and symmetries of spacetime may be uncovered by studying the first integrals of the geodesic equation describing a particle propagating on a curved background. The celebrated example is the discovery of a quadratic first integral for a massive particle moving in the Kerr spacetime \cite{c}, which preceded the construction of the second rank Killing tensor for the Kerr geometry \cite{wp}.

A spacetime may also admit an antisymmetric analogue of the Killing tensor, known in the literature as the Killing--Yano tensor.
In general, the Killing--Yano tensors may be used to construct the Killing tensors, but not every Killing tensor decomposes into a combination of the Killing--Yano tensors (see e.g. \cite{frolov}).

There are several reasons to be concerned about the Killing tensors and their antisymmetric analogues.
They give a clue for establishing the complete integrability of the geodesic equation and the complete separation of
variables for some important field equations in gravitational background \cite{c,carter} (a generalization of these results to higher dimensional Kerr--NUT--AdS black hole was performed in a series of recent works \cite{frolov1}--\cite{frolov3}).
They allow one to identify the spacetime in accord with the Petrov classification \cite{wp,kfk}. The Killing--Yano tensors underlie
the exotic supersymmetry \cite{grh}. It should also be mentioned that in the near horizon limit the isometry group of the extremal Kerr
metric is enhanced to include the conformal group $SO(2,1)$ \cite{bh} and the second rank Killing tensor becomes reducible \cite{g,go}.
For a massive particle moving on this background the Killing tensor governs the dynamics of the angular sector and specifies a reduced integrable system \cite{gn}. Other applications of the Killing tensors are discussed in a recent work \cite{s} where further references to the original literature can be found.

Although spacetimes admitting Killing tensors have been extensively investigated in the past,
no examples of irreducible Killing tensors of rank greater than four appear to be known.
In particular, in a recent work
\cite{ghkw} the Eisenhat lift \cite{eis} was applied to Goryachev--Chaplygin and Kovalevskaya's tops in order to construct new irreducible rank--3 and rank--4 Killing tensors. In \cite{gr} the results were extended to Goryachev--Chaplygin and Kovalevskaya's gyrostats and
the Brdi\v{c}ka-Eardley-Nappi-Witten plane--fronted wave with parallel rays (the $pp$-wave).

The purpose of this work is to construct an $(n+2)$--dimensional Lorentzian spacetime which admits irreducible Killing tensors of rank up to $n$.
This is achieved by applying the Eisenhart lift to the Calogero model \cite{cal}.

The Eisenhart lift \cite{eis}  is a specific embedding of a dynamical system with $n$ degrees of freedom into an $(n+2)$--dimensional Lorentzian spacetime such that the equations of motion of the original system are contained within the null geodesic equation.
It was originally introduced as a recipe of geometrization of Newtonian mechanics but it has fallen into oblivion soon.
After being rediscovered in \cite{duv,dgh} (where it was called the Bargmann space) the method proved to be very useful in
studying the issue of stability of mechanical systems and the description of non--relativistic symmetries (see e.g. \cite{pet,dhp,hor} and references therein).

The Calogero model \cite{cal} describes a set of identical particles on the real line interacting through
an inverse-square pair potential. It is one of only a few known many--body models which are integrable in classical domain
and exactly solvable after quantization. The range of physical applications of the Calogero model is impressive.
It includes fractional statistics \cite{frac}, gauge theory \cite{gor}, black hole physics \cite{gt},
the Witten-Dijkgraaf-Verlinde-Verlinde equation \cite{bgl,gala} and others.

The motivation for the present work is two--fold.
On the one hand, it is instructive to provide a description of the Calogero model in purely geometric terms. On the other hand,
the Calogero model is known to be maximally superintegrable \cite{w}, i.e. possessing the maximum allowed number of functionally independent integrals of motion. When considered within the Eisenhart framework, it gives rise to a Lorentzian spacetime with impressively large hidden symmetry.

The paper is organized as follows. In the next section we briefly review the structure of the integrals of motion and constants of the motion for the Calogero model. In Section 3 the Eisenhart lift is considered and a relation between the conserved quantities of an integrable system and symmetries of the spacetime is discussed. A criterion for the resulting Killing vectors and the Killing tensors to be conformal is formulated.
In Section 4 the Eisenhart lift is applied to the Calogero model and an
$(n+2)$--dimensional Lorentzian spacetime is constructed which admits irreducible Killing tensors of rank $3 \leq r \leq n$, where $n$ is the number of particles in the Calogero model. We summarize our results and discuss possible further developments in the concluding Section 5.

\vspace{0.5cm}

\noindent
{\bf 2. The Calogero model}\\

The Calogero model~\cite{cal} describes a set of $n$ identical particles on the real line interacting through
an inverse-square pair potential. Its Hamiltonian reads
\be\label{calogero}
H=\frac{1}{2} \sum_{i=1}^n p_i^2  +\sum_{i<j} \frac{g^2}{{(x_i-x_j)}^2},
\ee
where $g$ is a coupling constant. Throughout the paper we use the canonical Poisson brackets $\{x_i,p_j\}=\delta_{ij}$, $\{x_i,x_j\}=0$, $\{p_i,p_j\}=0$. A passage from the Hamiltonian formalism to the Lagrangian framework is established in the conventional way $p_i (t)=\dot x_i (t)$.
For simplicity we set the particle mass to unity.

That the model is integrable was first demonstrated by the method of isospectral deformation \cite{mos} (see also \cite{reg}).
The Lax matrix
\newpage
\be
L=\begin{pmatrix}
p_1 & \frac{i g}{(x_1-x_2)}& \dots &  \frac{i g}{(x_1-x_n)}\\
\frac{i g}{(x_2-x_1)} & p_2 & \dots & \frac{i g}{(x_2-x_n)}\\
\dots & \dots & \dots & \dots\\
\frac{i g}{(x_n-x_1)} & \frac{i g}{(x_n-x_2)} & \dots & p_n
\end{pmatrix},
\ee
determines $n$ independent integrals of motion
\be\label{L}
I_l=\frac{1}{l!} \tr L^l,
\ee
where $l=1,\dots,n$,
which are in involution \cite{mos,reg}. For our subsequent consideration it is important to stress that
$I_l$ is a polynomial of the $l$--th order in momenta. In particular, two lowest values reproduce the total momentum and the Hamiltonian, while
the next few integrals of motion read
\bea
&&
I_3=\frac{1}{3!} \left(\sum_{i=1}^n p_i^3+3 g^2\sum_{i<j} \frac{p_i+p_j}{{(x_i-x_j)}^2}\right),
\nonumber\\[2pt]
&&
I_4=\frac{1}{4!} \left(\sum_{i=1}^n p_i^4+4 g^2\sum_{i<j} \frac{p_i^2+p_j^2+p_i p_j}{{(x_i-x_j)}^2}+
2 g^4
\sum_{i<j} \frac{1}{{(x_i-x_j)}^4}+
\right.
\nonumber\\[2pt]
&&
\qquad \qquad
\left.
+4 g^4 \sum_{i \neq j, i \neq k, j<k} \frac{1}{{(x_i-x_j)}^2 {(x_i-x_k)}^2}\right),
\nonumber\\[2pt]
&&
I_5=\frac{1}{5!} \left( \sum_{i=1}^n p_i^5+5 g^2 \sum_{i<j} \frac{p_i^3+p_j^3+p_i^2 p_j+p_j^2 p_i}{{(x_i-x_j)}^2}+5 g^4 \sum_{i<j} \frac{p_i+p_j}{{(x_i-x_j)}^4}+
\right.
\nonumber\\[2pt]
&&
\qquad \qquad
\left.
+ 5 g^4 \sum_{i \neq j, i \neq k, j<k} \frac{2 p_i+p_j+p_k}{{(x_i-x_j)}^2 {(x_i-x_k)}^2}
\right).
\eea

A salient feature of the Calogero model is that $I_l$ can be used to generate extra constants of the motion
which, in principle, allow one to solve the equations of motion by purely algebraic means \cite{reg,w}.
Consider the following functions on the phase space\footnote{Within the method of isospectral deformation $M_l$
is linked to $\frac{1}{l!} \tr \left( Q L^{l-1}\right)$ with $Q_{ij}=x_i \delta_{ij}$ \cite{reg,w}.}
\be
M_l=\frac{1}{2l} \{ \sum_{i=1}^n x_i^2,I_l \},
\ee
where $l=1,\dots,n$.
Taking into account the Jacobi identity and the relations
\be\label{supl}
\frac 12 \{\sum_{i=1}^n x_i^2,H \}=\sum_{i=1}^n x_i p_i, \qquad \{\sum_{i=1}^n x_i p_i,I_l \}=l I_l,
\ee
one gets
\be\label{supl1}
\{M_l,H\}=I_l,
\ee
which implies that
\be
\tilde I_l=M_l-t I_l
\ee
are constants of the motion. Note that ${\tilde I}_l I_s-{\tilde I}_s I_l$ are conserved quantities which do not depend on time explicitly.
Together with $I_l$ they form $2n-1$ functionally independent integrals of motion of the Calogero model \cite{w}. For our subsequent consideration it proves convenient to allow conserved quantities which explicitly depend on time and to work in terms of a larger set which includes $I_l$ and $\tilde I_l$.

It is instructive to display a few lowest values of $M_l$ in explicit form
\bea
&&
M_1=\sum_{i=1}^n x_i, \qquad
M_2=\frac 12 \sum_{i=1}^n x_i p_i, \qquad
M_3=\frac{1}{3!} \left(\sum_{i=1}^n p_i^2 x_i+g^2\sum_{i<j} \frac{x_i+x_j}{{(x_i-x_j)}^2}\right),
\nonumber\\[2pt]
&&
M_4=\frac{1}{4!} \left(\sum_{i=1}^n p_i^3 x_i+g^2\sum_{i<j} \frac{2 p_i x_i+2 p_j x_j+x_i p_j+x_j p_i}{{(x_i-x_j)}^2}\right),
\nonumber\\[2pt]
&&
M_5=\frac{1}{5!} \left( \sum_{i=1}^n p_i^4 x_i+g^2 \sum_{i<j} \frac{3 p_i^2 x_i+3 p_j^2 x_j+p_i^2 x_j+p_j^2 x_i+2 p_i p_j x_i+2 p_j p_i x_j }{{(x_i-x_j)}^2}+
\right.
\nonumber\\[2pt]
&&
\qquad \qquad
\left. +g^4 \sum_{i<j} \frac{x_i+x_j}{{(x_i-x_j)}^4}
+ g^4 \sum_{i \neq j, i \neq k, j<k} \frac{2 x_i+x_j+x_k}{{(x_i-x_j)}^2 {(x_i-x_k)}^2}
\right).
\eea

It is straightforward to verify that the vectors $\frac{\partial I_l}{\partial \Gamma^A}$, $\frac{\partial \tilde I_l}{\partial \Gamma^A}$, where $\Gamma^A=(x_1,\dots,x_n,p_1,\dots,p_n,t)$, are linearly independent which means that $I_l$, $\tilde I_l$ are functionally independent.
Note that the algebra formed by $I_l$ and $\tilde I_l$ is non--linear. For sufficiently large values of $l$ and $s$ the brackets
$\{I_l,\tilde I_s\}$, $\{\tilde I_l, \tilde I_s \}$ yield rational functions which non--linearly depend on $I_l$ and $\tilde I_l$ with
$l=1,\dots,n$.

The Calogero model is conformal invariant. From (\ref{supl}), (\ref{supl1}) one deduces that
\be\label{Sc}
C=\frac 12 \sum_{i=1}^n x_i^2-t \sum_{i=1}^n x_i p_i+t^2 H
\ee
is a constant of the motion as well. The Poisson brackets of the triple $H$, $D=-\tilde I_2$, and $C$ reproduce the structure relations of $so(2,1)$
\be
\{H,D\}=H, \qquad \{H,C \}=2 D, \qquad \{D,C\}=C,
\ee
which is the conformal algebra in one dimension. It should be remembered, however, that $C$ is not functionally independent of the other constants of the motion. This can be verified by demonstrating that $\frac{\partial C}{\partial \Gamma^A}$ and $\frac{\partial I_l}{\partial \Gamma^A}$, $\frac{\partial \tilde I_l}{\partial \Gamma^A}$ are linearly dependent.
A simpler way is to notice that on--shell
\be
\frac 12 \sum_{i=1}^n x_i^2=\frac{1}{4H} {\left(\sum_{i=1}^n x_i p_i\right)}^2
\ee
up to an additive constant. The same conclusion is reached by looking at the realization of the Casimir element of $so(2,1)$ in the model
(\ref{calogero}) which implies that on--shell $C=D^2/H$.

For the discussion that follows it proves convenient to regard $n$ particles on the real line as one particle with $n$ degrees of freedom.
Lagrangian symmetry transformations which we consider in this work are of the form
\be
t'=t+\delta t (t), \qquad x'_i (t')=x_i(t)+\delta x_i(t,x(t)).
\ee
If the action functional $S=\int dt \mathcal{L}(x,\dot x)$ holds invariant under the transformation up to a total derivative, i.e. $\delta S=\int dt \Big( \frac{d F}{dt}\Big)$, then
the conserved quantity is derived from the expression
\be
\delta x_i \frac{\partial \mathcal{L}}{\partial \dot x_i}-
\delta t \Big( \dot x_i \frac{\partial \mathcal{L}}{\partial \dot x_i}-\mathcal{L} \Big)-F
\ee
by discarding the parameter of the transformation.

At the Lagrangian level the conserved charges $I_l$, $\tilde I_l$, with $l\leq 2$, can be linked to coordinate transformations in
$\mathbf{R}^{n} \times \mathbf{R}^{1}$.
Associated with $I_1$ and $I_2$ are translations of the spatial and temporal coordinates
\be
\delta x_i=\alpha, \qquad \delta t=\beta,
\ee
while $\tilde I_1$ and $\tilde I_2$ correspond to the boost
\be\label{boost}
\delta x_i=\gamma t,
\ee
and the dilatation
\be
\delta t=2 \lambda t, \qquad \delta x_i=\lambda x_i.
\ee
Here $\alpha$, $\beta$, $\gamma$ and $\lambda$ are infinitesimal parameters. It is instructive to display also the
special conformal transformation
\be\label{sc}
\delta t=\s t^2, \qquad \delta x_i= \s t x_i,
\ee
which is related to $C$ in (\ref{Sc}). Note that (\ref{boost}) and (\ref{sc}) leave the action functional of the Calogero model invariant up to a total derivative. Symmetry transformations corresponding to $I_l$, $\tilde I_l$, with $l>2$, involve velocities $\dot x_i (t)$ (see e.g. \cite{s}). Because it is problematic to link them to coordinate transformations in spacetime, within the geometric framework they are treated as
hidden symmetries.

Note that at this stage the geometry is that of the conventional Newtonian mechanics, i.e. $\mathbf{R}^{n} \times \mathbf{R}^{1}$, with the Euclidean metric defined on $\mathbf{R}^{n}$. In the next section, following Eisenhart, we shall introduce an extra coordinate and consider an $(n+2)$--dimensional Lorentzian spacetime a specific projection of which yields the configuration space of a particle with $n$ degrees of freedom extended by the temporal coordinate.

\vspace{0.5cm}

\noindent
{\bf 3. The Eisenhart lift}\\

The Eisenhart lift \cite{eis} (see also \cite{duv,dgh}) is an embedding of a dynamical system with $n$ degrees of freedom $x_1,\dots,x_n$ which is governed by a potential $U(x)$
\footnote{In general, the potential $U$ is allowed to depend on time explicitly. In this work we discuss only closed systems.}
into an $(n+2)$--dimensional Lorentzian spacetime parameterized by the coordinates $y^A=(x_1,\dots,x_n,t,s)$. It is defined
such that the equations of motion of the original system are contained within the null geodesic equations
\be\label{eisen}
\frac{d^2 y^A}{d \tau^2}+\Gamma^A_{BC} (y) \frac{d y^B}{d \tau} \frac{d y^C}{d \tau}=0, \qquad g_{AB}(y) \frac{d y^A}{d \tau} \frac{d y^B}{d \tau}=0,
\ee
specified by the metric
\be\label{metric}
d \tau^2=g_{AB}(y) d y^A d y^B=-2 U(x) d t^2+2 dt ds+\sum_{i=1}^n {(d x_i)}^2.
\ee
If the original particle was coupled to an external vector potential $A_i(x,t)$, then the metric would involve an extra contribution
$2 A_i(x,t) dt d x_i$. In what follows we use the notation in which the coordinates
$t$ and $s$ are designated explicitly, while $i=1,\dots,n$. Unless explicitly indicated otherwise, no summation over repeated indices is understood.

Taking into account the non--vanishing components of the Christoffel symbol
\be\label{cs}
\Gamma^i_{tt}=-\Gamma^s_{t i}=\partial_i U(x),
\ee
one can rewrite the geodesic equation in the form
\be\label{eqs}
\frac{d^2 x_i}{d t^2}+\partial_i U(x)=0, \qquad \frac{d t}{d \tau}=c_1, \qquad \frac{d s}{d t}-2 U(x)=c_2,
\ee
where $c_1$ and $c_2$ are arbitrary constants. The condition that the geodesic is null reads
\be\label{eqs1}
\frac 12 \sum_{i=1}^n {\left(\frac{d x_i}{d t}\right)}^2+\frac{d s}{d t}-U(x)=0.
\ee
The original dynamics is thus recovered by
implementing a null reduction along $s$ \cite{eis}. The relations above imply that $t$ can be interpreted as the temporal coordinate,
while $s$ is closely related to the action.

A salient feature of the Eisenhart framework is that the Killing vector $\xi=\frac{\partial}{\partial s}$ corresponding to
the isometry $s'=s+\nu$ of the metric (\ref{metric}) is null and covariantly constant. The Lorentzian spacetime (\ref{metric}) thus admits
a geodesic null congruence with vanishing expansion, shear and vorticity and belongs to the class of Kundt spacetimes.

Further specification occurs for harmonic functions $U(x)$. Given the Christoffel symbols (\ref{cs}), one can readily verify that
the only non--vanishing component of the Ricci tensor reads
\be
R_{tt}=\sum_{i=1}^n \partial_i \partial_i U(x)
\ee
and the scalar curvature vanishes. Thus, any harmonic function gives rise to the metric (\ref{metric}) which solves
the vacuum Einstein equations. Such solutions are known as the $pp$-waves. Note that the Calogero potential which we consider in this work does
not belong to this special class.

Let us now discuss how conserved charges of the original dynamical system are mapped into symmetries of the spacetime.
Recall that a totally symmetric tensor field $K_{A_1 \dots A_n}(y)$ is called a conformal Killing tensor if it obeys
the condition
\be\label{ckt}
\nabla_{(A_1} K_{A_2 \dots A_{n+1} )}=g_{(A_1 A_2} {\tilde K}_{A_3 \dots A_{n+1})},
\ee
where the explicit form of the tensor ${\tilde K}_{A_1 \dots A_{n-1}}(y)$ is found by taking the trace of the both sides of (\ref{ckt}).
If the components ${\tilde K}_{A_1 \dots A_{n-1}}(y)$ happen to vanish, one has the usual Killing tensor.
Because within the Eisenhart framework the geodesic is null, in general, a conserved charge of a dynamical systems yields
a conformal Killing tensor. To be more specific, in view of (\ref{eqs}), a multiplication of a conserved charge which is a polynomial in momenta $p_i=\frac{d x_i}{d t}$ of degree $l$ by ${\left(\frac{d t}{d \tau}\right)}^l$ yields an expression of the form
$K_{A_1 \dots A_l} (y) \frac{d y^{A_1}}{d \tau}\dots \frac{d y^{A_l}}{d \tau}$ from which the Killing tensor $K_{A_1 \dots A_l} (y)$ is derived.
A criterion for the resulting Killing tensor to be conformal is prompted by the dynamical system itself. If the derivative of the integral of motion with respect to time leads to the expression which appears in the left hand side of (\ref{eqs1}) (the condition that the geodesic is null) then the resulting Killing tensor will be conformal. In particular,
for the Calogero model which we consider in the next section none of the Killing tensors proves to be conformal, while among six Killing vectors only two are conformal.

\vspace{0.5cm}

\noindent
{\bf 4. Higher rank Killing tensors and the Calogero model}\\

Let us see in more detail how the method outlined in the preceding section works for the Calogero model. Putting into the left column the first integrals of the geodesic equation and into the right column the corresponding Killing vectors, one finds
\begin{align}\label{int}
&
\frac{d t}{d \tau}, && \frac{\partial}{\partial s}
\nonumber
\\[2pt]
&
\frac{d s}{d \tau}-2 U(x) \frac{d t}{d \tau}, && \frac{\partial}{\partial t}
\nonumber
\\[2pt]
&
\sum_{i=1}^n \frac{d x_i}{d \tau}, && \sum_{i=1}^n \frac{\partial}{\partial x_i}
\nonumber
\end{align}
\begin{align}
&
\sum_{i=1}^n x_i  \frac{d t}{d \tau}-t \sum_{i=1}^n \frac{d x_i}{d \tau}, && t \sum_{i=1}^n \frac{\partial}{\partial x_i}-\sum_{i=1}^n x_i \frac{\partial}{\partial s}
\nonumber
\\[2pt]
&
\frac 12 \sum_{i=1}^n x_i  \frac{d x_i}{d \tau} + t \left(\frac{d s}{d \tau}-2 U(x) \frac{d t}{d \tau}   \right),  && 2 t \frac{\partial}{\partial t}+\sum_{i=1}^n x_i \frac{\partial}{\partial x_i}
\nonumber
\\[2pt]
&
\frac 12 \sum_{i=1}^n x_i^2 \frac{d t}{d \tau}-t \sum_{i=1}^n x_i  \frac{d x_i}{d \tau}-t^2 \left(\frac{d s}{d \tau}-2 U(x) \frac{d t}{d \tau}   \right),   && t^2 \frac{\partial}{\partial t}-\frac 12 \sum_{i=1}^n x_i^2 \frac{\partial}{\partial s}+t \sum_{i=1}^n x_i \frac{\partial}{\partial x_i}.
\end{align}
Here we used the fact that the Calogero potential
$U(x)=\sum_{i<j} \frac{g^2}{{(x_i-x_j)}^2}$ is translation and conformal invariant
\be
\sum_{i=1}^n \partial_i U(x)=0, \qquad \sum_{i=1}^n x_i \partial_i U(x)=-2 U(x).
\ee

Taking into account (\ref{eqs1}), one can readily verify that the left column in (\ref{int}) reproduces
$I_1$, $\tilde I_1$, $I_2$, $\tilde I_2$ in the previous section. Besides, $t$ can be identified with the temporal coordinate and the dynamics of $s$ is fixed provided the evolution of $x_i$ is known. Note that within the geometric framework the boost (\ref{boost}) and the special conformal transformation (\ref{sc}) are extended by a transformation of the variable $s$. This is a manifestation of the fact that the action functional of the Calogero model holds invariant under (\ref{boost}) and (\ref{sc}) up to a total derivative. In accord with the criterion we formulated at the end of the preceding section, the first four lines entering the right column in (\ref{int}) give the Killing vector fields, while the last two
determine the conformal Killing vectors.

Hidden symmetries of the metric (\ref{metric}) are derived from $I_l$, $\tilde I_l$, with $l>2$. For example, $I_3$, $\tilde I_3$ yield a couple of the third rank Killing tensors  $K^{(3)}_{ABC}$ and ${\tilde K}^{(3)}_{ABC}$
\bea
&&
K^{(3)}_{iii}=1, \qquad K^{(3)}_{tti}=\sum_{j=1, j \ne i}^n \frac{g^2}{{(x_i-x_j)}^2};
\\[2pt]
&&
{\tilde K}^{(3)}_{iii}=-t, \quad {\tilde K}^{(3)}_{tti}=-t \sum_{j=1, j\ne i}^n \frac{g^2}{{(x_i-x_j)}^2}, \quad {\tilde K}^{(3)}_{tii}=\frac 13 x_i,
\quad {\tilde K}^{(3)}_{ttt}= g^2 \sum_{i<j} \frac{x_i+x_j}{{(x_i-x_j)}^2},
\nonumber
\eea
$I_4$, $\tilde I_4$ give rise to the fourth rank Killing tensors  $K^{(4)}_{ABCD}$ and ${\tilde K}^{(4)}_{ABCD}$
\bea
&&
K^{(4)}_{iiii}=1,\quad
K^{(4)}_{tttt}=
2
\sum_{i<j} \frac{g^4}{{(x_i-x_j)}^4}+
4  \sum_{i \neq j, i \neq k, j<k} \frac{g^4}{{(x_i-x_j)}^2 {(x_i-x_k)}^2},
\nonumber\\[2pt]
&&
K^{(4)}_{ttii}=\frac 23 \sum_{j=1, j\ne i}^n \frac{g^2}{{(x_i-x_j)}^2}, \quad K^{(4)}_{ttij}=\frac 13 \frac{g^2}{{(x_i-x_j)}^2};
\nonumber\\[2pt]
&&
{\tilde K}^{(4)}_{iiii}=-t,\quad
{\tilde K}^{(4)}_{tttt}=
-2 t
\sum_{i<j} \frac{g^4}{{(x_i-x_j)}^4}-
4 t  \sum_{i \neq j, i \neq k, j<k} \frac{g^4}{{(x_i-x_j)}^2 {(x_i-x_k)}^2},
\nonumber
\eea
\bea
&&
{\tilde K}^{(4)}_{ttii}=- \frac 23 t \sum_{j=1, j\ne i}^n \frac{g^2}{{(x_i-x_j)}^2}, \quad {\tilde K}^{(4)}_{ttij}=-\frac 13 t \frac{g^2}{{(x_i-x_j)}^2},
\quad {\tilde K}^{(4)}_{tiii}=\frac 14 x_i,
\nonumber\\[2pt]
&&
{\tilde K}^{(4)}_{ttti}=\frac 14 g^2 \sum_{j=1, j\ne i}^n \frac{2 x_i+x_j}{{(x_i-x_j)}^2},
\eea
while $I_5$, $\tilde I_5$ produce Killing tensors  $K^{(5)}_{ABCDE}$ and ${\tilde K}^{(5)}_{ABCDE}$ of the fifth rank
\bea
&&
K^{(5)}_{iiiii}=1, \qquad \quad
K^{(5)}_{ttiii}=\frac 12 \sum_{j=1, j \ne i}^n \frac{g^2}{{(x_i-x_j)}^2}, \qquad \quad K^{(5)}_{ttiij}=\frac 16 \frac{g^2}{{(x_i-x_j)}^2},
\nonumber\\[2pt]
&&
K^{(5)}_{tttti}=\sum_{j=1, j \ne i}^n \frac{g^4}{{(x_i-x_j)}^4}+2 \sum_{j,k=1, j<k}^n \frac{g^4}{{(x_i-x_j)}^2 {(x_i-x_k)}^2}+
\sum_{j,k=1, i \ne k}^n \frac{g^4}{{(x_i-x_j)}^2 {(x_k-x_j)}^2};
\nonumber\\[2pt]
&&
{\tilde K}^{(5)}_{iiiii}=-t, \qquad
{\tilde K}^{(5)}_{ttiii}=-\frac 12 t \sum_{j=1, j \ne i}^n \frac{g^2}{{(x_i-x_j)}^2}, \qquad {\tilde K}^{(5)}_{ttiij}=-\frac 16 t \frac{g^2}{{(x_i-x_j)}^2},
\eea
\bea
&&
{\tilde K}^{(5)}_{tttti}=-\sum_{j=1, j \ne i}^n \frac{t g^4}{{(x_i-x_j)}^4}- \sum_{j,k=1, j<k}^n \frac{2 t g^4}{{(x_i-x_j)}^2 {(x_i-x_k)}^2}-
\sum_{j,k=1, i \ne k}^n \frac{t g^4}{{(x_i-x_j)}^2 {(x_k-x_j)}^2},
\nonumber\\[2pt]
&&
{\tilde K}^{(5)}_{tiiii}=\frac 15 x_i, \qquad ~ {\tilde K}^{(5)}_{tttij}=\frac{g^2}{10} \frac{(x_i+x_j)}{{(x_i-x_j)}^2}, \qquad \quad
{\tilde K}^{(5)}_{tttii}=\frac{g^2}{10} \sum_{j=1, j\ne i}^n \frac{3 x_i+x_j}{{(x_i-x_j)}^2},
\nonumber\\[2pt]
&&
{\tilde K}^{(5)}_{ttttt}=\sum_{i<j}^n \frac{g^4 (x_i+x_j)}{{(x_i-x_j)}^4}+\sum_{i,j,k=1, i<k}^n \frac{g^4 (x_i+x_k)}{{(x_i-x_j)}^2 {(x_k-x_j)}^2}
+\sum_{i,j,k=1, j<k}^n \frac{2 g^4 x_i}{{(x_i-x_j)}^2 {(x_i-x_k)}^2}.
\nonumber
\eea
Other Killing tensors are built likewise. By construction, they are irreducible. None of them proves to be conformal.

\vspace{0.5cm}


\noindent
{\bf 5. Conclusion}\\

To summarize, in this work we have constructed an $(n+2)$--dimensional Lorentzian spacetime which admits irreducible Killing tensors of rank $3 \leq r \leq n$. This was achieved by applying the Eisenhart lift to the Calogero model. Because within the Eisenhart framework
the equations of motion of a dynamical system are embedded into the null geodesic equation, the Killing vectors and the Killing tensors associated to the integrals of motion and constants of the motion of the original dynamical system are allowed to be conformal. In particular, the spacetime constructed in this work admits conformal Killing vectors but no conformal Killing tensor. It would be interesting to construct a spacetime with  irreducible higher rank conformal Killing tensors in a similar fashion. Spacetimes corresponding to integrable generalizations of the Calogero model, such as
the Calogero model in a harmonic trap or its extension by spin degrees of freedom, are also worthy of study.

\vspace{0.5cm}

\noindent{\bf Acknowledgements}\\

\noindent
This work was supported by the Dynasty Foundation, RF Federal Program "Kadry" under
the contracts 16.740.11.0469, P691, RFBR grant 11-02-90445, and LSS grant 224.2012.2.

\end{document}